\documentclass[aps,prd,showpacs,twocolumn,notitlepage,
superscriptaddress]{revtex4-1}
\usepackage{graphicx}
\usepackage{dcolumn}

\usepackage{amsmath}
\usepackage{amssymb}
\usepackage{bm}
\usepackage{color}
\usepackage[normalem]{ulem}
\usepackage[dvipsnames]{xcolor}
\usepackage{hyperref}
\hypersetup{
colorlinks=true, % false: boxed links; true: colored links
linkcolor=blue,  % color of internal links
citecolor=cyan,  % color of links to bibliography
}

\begin{document}
\title{Anisotropic Tolman VII solution by gravitational decoupling}

\author{Sudipta Hensh}
\email{f170656@fpf.slu.cz, sudiptahensh2009@gmail.com}
\affiliation{Institute of Physics and Research Centre of Theoretical Physics and Astrophysics,Faculty of Philosophy \& Science, Silesian University in Opava,
Bezru\v{c}ovo n\'{a}m\v{e}st\'{i} 13, CZ-74601 Opava, Czech Republic}

%
%\author{Jorge Ovalle}
%\email{jovalle@usb.ve}
%\affiliation{Departamento de F´ısica, Universidad Sim´on Bol´ıvar, AP 89000, Caracas 1080A, Venezuela }
%\affiliation{Faculty of Philosophy \& Science, Institute of Physics, Silesian University in Opava,
%Bezru\v{c}ovo n\'{a}m\v{e}st\'{i} 13, CZ-74601 Opava, Czech Republic }

\author{Zden\v{e}k Stuchl\'{i}k}
\email{zdenek.stuchlik@fpf.slu.cz}
\affiliation{Institute of Physics and Research Centre of Theoretical Physics and Astrophysics,Faculty of Philosophy \& Science, Silesian University in Opava,
Bezru\v{c}ovo n\'{a}m\v{e}st\'{i} 13, CZ-74601 Opava, Czech Republic}

\date{\today}

\begin{abstract}
Using the gravitational decoupling by the minimal geometric deformation approach, we build an anisotropic version of the well-known Tolman VII solution, determining an exact and physically acceptable interior two-fluid solution that can represent behavior of compact objects. Comparison of the effective density and density of the perfect fluid is demonstrated explicitly. We show that the radial and tangential pressure are different in magnitude giving thus the anisotropy of the modified Tolman VII solution. The dependence of the anisotropy on the coupling constant is also shown.
\end{abstract}

\maketitle

%

%%%%%%%%%%%%%%%%%%%%%%%%%%%%%%%%%%%%%%%%%%%%%%%%%%%%%%%%%%%%%%%%%%%%%%%%%%%%%%%%
\section{Introduction}\label{intro}
Einstein's gravitational field equations are partial nonlinear differential equations -- their solution is very difficult with exception of some simplified situations. Immediately after Einstein introduced the general relativity(GR), K. Schwarzschild solved the vacuum Einstein equations~\cite{schwarzschild1916gravitationsfeld} describing exterior of a spherically symmetric and static sphere. The simple internal spherically symmetric solution with special uniform distribution of matter has been found by Schwarzschild~\cite{schwarzschild1916gravitationsfeld}, and generalized for spacetimes with non zero cosmological constant in~\cite{stuchlik2000spherically,bohmer2004eleven}. Some important internal solutions of the Einstein equations were found by R. Tolman for perfect fluid spheres with fluid described by polytropic equations of state~\cite{tolman1939static}. The polytropes in spacetimes with non-zero cosmological constant were extensively discussed in \cite{stuchlik2000spherically,stuchlik2016general}. Interesting properties of the spherically symmetric polytropes were discussed in \cite{stuchlik2016general,novotny2017polytropic,hod2018lower,hod2018analytic,stuchlik2017gravitational}.

Anisotropic pressure in stellar distribution implies unequal radial and tangential pressure. The possible reasons for anisotropies in fluid pressure are presence of mixture of different fluids, different kinds of phase transitions \cite{sokolov1980phase}, viscosity, rotation, magnetic field, superfluid \cite{kippenhahn1990stellar} or existence of a solid core.  From our general notion, it is understandable that anisotropic solutions could represent a realistic description of astrophysical interest. Anisotropy of pressure in a perfect fluid sphere was first  described by G. Lemaitre \cite{lemaitre1933universe}. In 1974, Bower and Liang \cite{bowers1974anisotropic} described importance of locally anisotropic equations of state for relativistic spheres. Ruderman has shown in his significant work that nuclear matter may be anisotropic in very high density regions ($\rho>10^{17} \mathrm{kg/m^3}$) \cite{ruderman1972pulsars}. A few articles devoted to generating anisotropic solutions are available in literature \cite{thirukkanesh2018anisotropic,herrera2008all,chaisi2006new}.

There are eight perfect fluid solutions of the Einstein field equations presented by Tolman \cite{tolman1939static}. Among these solutions, the Tolman IV and the Tolman VII solutions are physically interesting and could depict some neutron star configurations. For detailed analysis of the Tolman VII solution see~\cite{sotani2018compactness,kiess2017voids,bhar2017compact,singh2016well,
raghoonundun2016exact,raghoonundun2016self,raghoonundun2016geometrical,
bhar2015relativistic,raghoonundun2015possible,
papazoglou2016r,kiess2012exact,neary2001r,neary2001tolman}. 

Ovalle introduced an anisotropic version of the isotropic Tolman IV solution in \cite{ovalle2018anisotropic}, by using the so called minimal geometric deformation~(MGD) method developed by himself~\cite{ovalle2017decoupling,ovalle2018anisotropic}. Our objective in the present paper is to generate a new anisotropic solution from the isotropic Tolman VII solution \citep{tolman1939static}, by using the Ovalle MGD method.

The MGD method is the first simple, systematic and direct method of decoupling gravitational sources in GR. Initially, MGD method was proposed \cite{ovalle2008searching,ovalle2010braneworld} in 
the context of the Randall-Sundrum braneworld model~\cite{randall1999alternative,randall1999large} --references for earlier works on the MGD method,~
\cite{ovalle2009nonuniform,
ovalle2010schwarzschild,
casadio2012brane,ovalle2013tolman,
ovalle2013role,casadio2014black}, and for some  recent applications see~
\cite{ovalle2015brane,casadio2015classical,cavalcanti2016strong,casadio2016stability,ovalle2017minimal,
da2017dark,da2017black,fernandes2018gregory,casadio2017gup,ovalle2018anisotropic,heras2018using,
fernandes2018extended,estrada2018new,contreras2018minimal,morales2018charged,gabbanelli2018gravitational,
sharif2018gravitational,sharif2018gravitationalsph,graterol2018new,ovalle2018black,
contreras2019gravitational,contreras2019general,maurya2019generalized,fernandes2019extended,
contreras2019gravitational,ovalle2019decoupling,ovalle2018einstein,sharif2018gravitationalde}).  The notable feature of the MGD method is that it preserves the spherical symmetry, as well as the physical acceptability. MGD method thus opens up a new window to search for physically acceptable anisotropic solutions.  

We review first the formulation of decoupling the Einstein field equations for different gravitational sources. We have to solve
\begin{equation}\label{eq:1}
\hat{G}_{\mu\nu}=-k^2{\hat{T}_{\mu\nu}};\quad \textrm{to find the metric}\  \hat{g}_{\mu\nu},
\end{equation}
and
\begin{equation}\label{eq:2}
{\tilde{G}_{\mu\nu}}=-k^2{\tilde{T}_{\mu\nu}};\quad \textrm{to find the metric}\  \tilde{g}_{\mu\nu},
\end{equation}
instead of looking for solution for the total energy momentum tensor $T_{\mu\nu}=\hat{T}_{\mu\nu}+\tilde{T}_{\mu\nu}$,
\begin{equation}\label{eq:3}
{G_{\mu\nu}}=-k^2{T_{\mu\nu}};\quad \textrm{to find the metric}\  g_{\mu\nu},
\end{equation}
where the constant $k^2=8\pi G/c^4$, and in the geometric units~($c=G=1$) there is $k^2=8\pi$. $T_{\mu\nu}$ is the total energy momentum tensor, $\hat{T}_{\mu\nu}$ is the energy momentum tensor for one gravitational source and $\tilde{T}_{\mu\nu}$ is the energy momentum tensor for another gravitational source. After solving Eqs.~(\ref{eq:1}) and (\ref{eq:2}), we can find metric $g_{\mu\nu}$ by combining $\hat{g}_{\mu\nu}$ and $\tilde{g}_{\mu\nu}$.

Generally, we can extend this formalism for any number of gravitational sources. In that case we have to solve Einstein's field equations for each source term, and combine the separately found metrics in order to get the metric related to the total energy momentum tensor. The number of physically acceptable solutions is not large. Delgaty and Lake examined physically acceptability of 127 known isotropic solutions~ \cite{delgaty1998physical}. They found that only 16 of them has physical relevance. 

The paper is organized as follows. In Section \ref{fieldeq} we discuss decoupling of Einstein's field equations. In Section \ref{mgd} we introduce the MGD method. In Section \ref{matching} we study the condition for matching the interior solution to the exterior one. Section \ref{perfect} is devoted to description of  the Tolman~VII perfect fluid solution. In Section \ref{aniso} we derive the new anisotropic solution and demonstrate its interesting properties. In Section \ref{conclu} conclusions are presented. 

%%%%%%%%%%%%%%%%%%%%%%%%%%%%%%%%%%%%%%%%%%%%%%%%%%%%%%%%%%%%%%%%%%%%%%%%%%%%%%%%

\section{Decoupling of Einstein's field equations}\label{fieldeq}

We shortly review the method of the gravitational decoupling developed by Ovalle \cite{ovalle2017decoupling}.
In the framework of the decoupling method, Einstein's field equations are expressed in the form
\begin{equation}\label{eq:4}
R_{\mu\nu}-\frac{1}{2}Rg_{\mu\nu}=-k^2T_{\mu\nu}^{\mathrm{(tot)}} \ ,
\end{equation}
where,
\begin{equation}\label{eq:5}
{T}_{\mu\nu}^{\mathrm{(tot)}}={T}_{\mu\nu}^{(\mathrm{perfect fluid)}}+{T}_{\mu\nu}^{\mathrm{(other source)}} \ .
\end{equation}
The energy momentum tensor for perfect fluid reads
\begin{equation}\label{eq:6}
{T}_{\mu\nu}^{(\mathrm{perfect fluid)}}=\left(\rho + p\right)u_{\mu}u_{\nu}-pg_{\mu\nu} \ ,
\end{equation}
where, $\rho$,$\;$ $p$ and $u_{\mu}$ are density, pressure and four velocity of perfect fluid, respectively.
We consider the contribution from other gravitational source($\theta_{\mu\nu}$) modified by an intensity parameter~$\alpha$
\begin{equation}\label{eq:7}
{T}_{\mu\nu}^{\mathrm{(other source)}}=\alpha \theta_{\mu\nu} \ .
\end{equation}
The total energy momentum tensor for perfect fluid coupled with another gravitational source causing anisotropy in the self gravitating system then reads 
\begin{equation}\label{eq:8}
{T}_{\mu\nu}^{\mathrm{(tot)}}=\left(\rho + p\right)u_{\mu}u_{\nu}-pg_{\mu\nu}+\alpha \theta_{\mu\nu} \ .
\end{equation}
The term $\theta_{\mu\nu}$ in Eq.~(\ref{eq:7}) stands for any source like scalar, vector, or tensor field, causing anisotropies in the fluid. As a consequence of Bianchi identity the conservation law holds 
\begin{equation}\label{eq:9}
\nabla_{\nu}{T}^{\mathrm{(tot)} \mu\nu} = 0 \ .
\end{equation}
The line element of spherically symmetric spacetime in Schwarzschild coordinates $\left(t,r,\theta,\phi\right)$ takes the form  
\begin{equation}\label{eq:10}
ds^2 = e^{\nu(r)}  dt^2 - e^{\lambda(r)} dr^2 - r^2(d\theta^2 + sin^2\theta  d\phi^2) \ ,
\end{equation}
where, $\nu(r)$ and $\lambda(r)$ are functions of the radial coordinate~($r$) which ranges from the compact object centre~($r=0$) to its surface~($r=R$). Four-velocity of the static fluid is given by $u^\mu=e^{-\nu/2} \delta_{0}^{\mu}$, at radii $0 \leq r \leq R$. The general metric given by Eq.~(\ref{eq:10}) obeys the Einstein field equations having the following form 
\begin{equation}\label{eq:11}
-k^2\left(\rho+ \alpha \theta_{0}^{0}\right)=-\frac{1}{r^2}+e^{-\lambda}\left(\frac{1}{r^2}-\frac{\lambda'}{r}\right) \ ,
\end{equation}
\begin{equation}\label{eq:12}
-k^2\left(-p+ \alpha \theta_{1}^{1}\right)=-\frac{1}{r^2}+e^{-\lambda}\left(\frac{1}{r^2}-\frac{\nu'}{r}\right) \ ,
\end{equation}
\begin{equation}\label{eq:13}
-k^2\left(-p+ \alpha \theta_{2}^{2}\right)=\frac{1}{4} e^{-\lambda}\left(2\nu''+\nu'^2-\lambda'\nu'+2\frac{\nu'-\lambda'}{r}\right) \ .
\end{equation}
The conservation Eq.~(\ref{eq:9}), which is linear combination of Eqs.~(\ref{eq:11}), (\ref{eq:12}) and (\ref{eq:13}), reads
\begin{equation}\label{eq:14}
-p'-\frac{\nu'}{2}\left(\rho+p\right)+\alpha \left(\theta_{1}^{1}\right)'-\frac{\nu'}{2}\alpha\left(\theta_{0}^{0}-\theta_{1}^{1}\right)-\frac{2}{r}\alpha\left(\theta_{2}^{2}-\theta_{1}^{1}\right)=0 \ .
\end{equation}
Here $'$ denotes differentiation of the function with respect to $r$. We can easily identify from Eqs.~(\ref{eq:11}), (\ref{eq:12}) and (\ref{eq:13}) the effective density
\begin{equation}\label{eq:15}
\rho_\mathrm{eff}\mathrm{(radial)}=\rho+\alpha \theta_{0}^{0} \ ,
\end{equation}
the effective isotropic pressure 
\begin{equation}\label{eq:16}
p_{r}=p-\alpha \theta_{1}^{1} \ ,
\end{equation}
and the effective tangential pressure
\begin{equation}\label{eq:17}
p_{t}=p-\alpha \theta_{2}^{2} \ .
\end{equation}
It is evident that the $\theta_{\mu\nu}$ source introduces an anisotropy into perfect fluid which is given by
\begin{equation}\label{eq:18}
\pi=p_{t}-p_{r}=\alpha (\theta_{2}^{2}-\theta_{1}^{1}) \ .
\end{equation}
Now, we have five unknowns, namely the metric coefficients $\nu(r)$, $\lambda(r)$, the effective density~($\rho_\mathrm{eff}$), the effective radial pressure~($p_{r}$), and the effective tangential pressure~($p_{t}$). To solve the Einstein equations to obtain these functions, we have to proceed using the MGD method.

%%%%%%%%%%%%%%%%%%%%%%%%%%%%%%%%%%%%%%%%%%%%%%%%%%%%%%%%%%%%%%%%%%%%%%%%%%%%%%%%%%

\section{Minimal geometric deformation}\label{mgd}

MGD method is a very strong technique to decouple Einstein's field equations for different source terms. This method has been developed in a simple and elegant way recently~\cite{ovalle2017decoupling}, we shortly review this method.
We consider a perfect fluid solution of energy density $\rho$ and pressure $p$ that is described by the metric coefficients $\left(e^\epsilon,e^\gamma\right)$. The line element of the corresponding spherically symmetric solution reads
\begin{equation}\label{eq:19}
ds^2 = e^{\epsilon(r)}  dt^2 - e^{\gamma(r)} dr^2 - r^2(d\theta^2 + sin^2\theta  d\phi^2) \ ,
\end{equation}
where the radial metric coefficient takes the form 
\begin{equation}\label{eq:20}
e^{\gamma(r)}=\left(1-\frac{2m(r)}{r}\right)^{-1} \ ,
\end{equation}
with $m(r)$ being so called GR mass function. We introduce the MGD transformations in the form \cite{ovalle2017decoupling} 
\begin{equation}\label{eq:21}
e^\epsilon \mapsto   e^\nu=e^{\epsilon+\alpha g} \ ,
\end{equation}
\begin{equation}\label{eq:22}
e^{-\gamma} \mapsto   e^{-\lambda}=e^{-\gamma}+\alpha f \ ,
\end{equation}
where $g$ and $f$ are deformations of temporal  and radial metric coefficients respectively, arising as an effect of introduction of the anisotropy. The minimal geometric deformation is given by the conditions
\begin{equation}\label{eq:23}
g \mapsto 0 \ ,
\end{equation}
and 
\begin{equation}\label{eq:24}
f \mapsto f^* \ .
\end{equation}
Then temporal component of the metric $e^\nu$ remains unchanged, while the additional gravitational source~($\theta_{\mu\nu}$) causes a deformation of the radial component according to Eq.~(\ref{eq:22}). If we incorporate the deformed metric into Eqs. (\ref{eq:11}), (\ref{eq:12}) and (\ref{eq:13}), we see that each equation can be decomposed into two equations -- one holds for the perfect fluid, the other one is involving~$\theta_{\mu\nu}$. 
The set of equations for perfect fluid~($\alpha=0$) is fully determined by the metric coefficient $\epsilon(r)=\nu(r)$ and takes the form 
\begin{equation}\label{eq:25}
k^2 \rho=\frac{1}{r^2}-e^{-\gamma}\left(\frac{1}{r^2}+\frac{\gamma'}{r}\right) \ ,
\end{equation}
\begin{equation}\label{eq:26}
k^2 p=-\frac{1}{r^2}+e^{-\gamma}\left(\frac{1}{r^2}+\frac{\nu'}{r}\right) \ ,
\end{equation}
\begin{equation}\label{eq:27}
k^2 p=\frac{1}{4}e^{-\gamma}\left(2\nu''+\nu'^2+\frac{2\nu'}{r}\right)+\frac{1}{4}\gamma' e^{-\gamma}\left(\nu'+\frac{2}{r}\right) \ .
\end{equation}
The conservation equation takes the form 
\begin{equation}\label{eq:28}
p'=-\frac{\nu'}{2}(\rho+p) \ .
\end{equation}
The equations involving the $\theta_{\mu\nu}$ term read
\begin{equation}\label{eq:29}
k^2 \theta_{0}^{0}=-\frac{f^*}{r^2}-\frac{f^{*'}}{r} \ ,
\end{equation}
\begin{equation}\label{eq:30}
k^2 \theta_{1}^{1}=- f^*\left(\frac{1}{r^2}+\frac{\nu'}{r}\right) \ ,
\end{equation}
\begin{equation}\label{eq:31}
k^2 \theta_{2}^{2}=-\frac{f^*}{4}\left(2\nu^{''}+\nu'^2+\frac{2\nu'}{r}\right)-\frac{f^{*'}}{4}\left(\nu'+\frac{2}{r}\right) \ .
\end{equation}
The conservation equation for the $\theta_{\mu\nu}$ term takes the form 
\begin{equation}\label{eq:32}
(\theta_{1}^{1})'-\frac{\nu'}{2}(\theta_{0}^{0}-\theta_{1}^{1})-\frac{2}{r}(\theta_{2}^{2}-\theta_{1}^{1})=0 \ ,
\end{equation}
and it is a linear combination of Eqs. (\ref{eq:29}), (\ref{eq:30}) and (\ref{eq:31}).

We see that Einstein's equations~(\ref{eq:11}), (\ref{eq:12}) and (\ref{eq:13}) are decoupled by deforming the radial metric component according to Eq. (\ref{eq:21}). The conservation Eqs.~(\ref{eq:28}) and (\ref{eq:32}) for the MGD solution have to be satisfied simultaneously with the general conservation law given by Eq.~(\ref{eq:9}). We conclude that both systems, perfect fluid and other gravitational source, conserve independently, i.e., these two systems cannot exchange energy-momentum, and their interaction is solely gravitational \cite{ovalle2017decoupling}.
%%%%%%%%%%%%%%%%%%%%%%%%%%%%%%%%%%%%%%%%%%%%%%%%%%%%%%%%%%%%%%%%%%%%%%%%%%%%%%%%%

\section{Matching condition}\label{matching}

\begin{figure*}[t!]
\begin{center}
\includegraphics{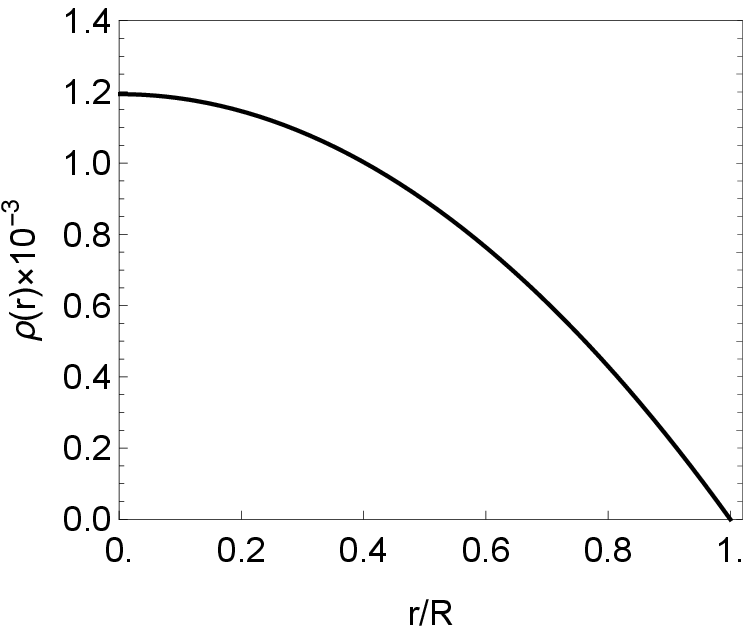}
\hspace{0.7cm}
\includegraphics{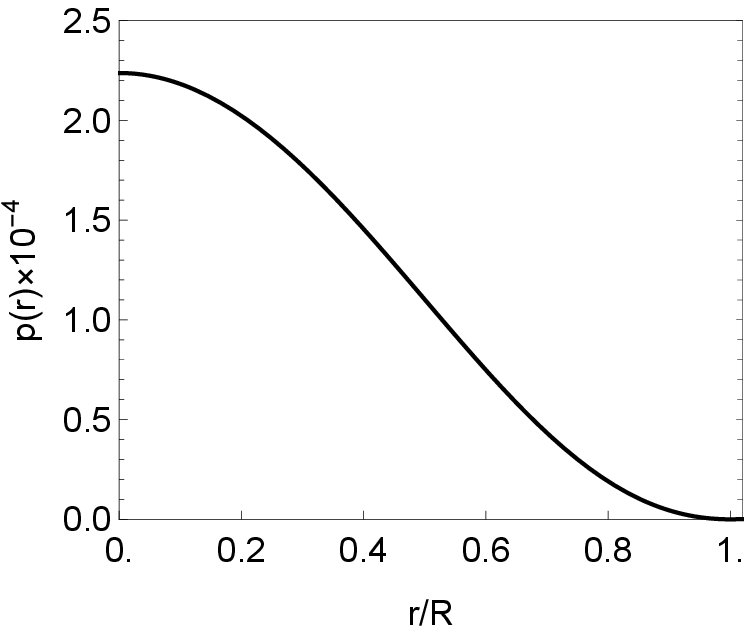}
\end{center}
\caption{This pictures depict the radial dependance of density and pressure for Tolman VII perfect fluid solution. The pictures are drawn for $M_0=2$ and $R=10$. \label{perquan}}
\end{figure*}

The matching of the interior~($r<R$) and exterior~($r>R$) solutions of any mass distribution at the boundary of the interior~($r=R$), has to be considered very carefully in the case of mixed sources treated in the framework of the MGD method~\cite{ovalle2017decoupling}. The interior metric of our consideration is given by Eq.~(\ref{eq:19}) along with Eqs~(\ref{eq:20}), (\ref{eq:21}), (\ref{eq:22})~and~(\ref{eq:23}), and its line element can be written as 
\begin{equation}\label{eq:33}
ds^2=e^{\nu^-(r)} dt^2-\left(1-\frac{2\tilde{m}}{r}\right)^{-1} dr^2- r^2(d\theta^2 + sin^2\theta  d\phi^2) \ ,
\end{equation}
where the interior mass function reads 
\begin{equation}\label{eq:34}
\tilde{m}=m(r)-\frac{r}{2} \alpha f^*(r) \ ,
\end{equation} 
$m(r)$ is the standard GR mass function in Eq.(\ref{eq:20}), and the function $f^*$ has to be calculated later. We assume there is no matter outside, i.e., $\rho^{+}=p^{+}=0$. In general, additional gravitational source $\theta_{\mu\nu}$ could influence the exterior geometry, and in such a case the general exterior metric can be written as 
\begin{equation}\label{eq:35}
ds^2=e^{\nu^+(r)} dt^2-e^{\lambda^+(r)} dr^2-r^2(d\theta^2 + sin^2\theta  d\phi^2) \ ,
\end{equation}
where the functional form of $\nu^+(r)$ and $\lambda^+(r)$ can be determined by solving Einstein's equations for the exterior geometry 
\begin{equation}\label{eq:36}
R_{\mu\nu}-\frac{1}{2} R g_{\mu\nu}=-k^2 \alpha \theta_{\mu\nu} \ .
\end{equation}
Continuity of the first fundamental form implies
\begin{equation}\label{eq:37}
ds^2\vert_{r \rightarrow R^+}-ds^2\vert_{r \rightarrow R^-}=0 \ .
\end{equation}
From above we obtain equations
\begin{equation}\label{eq:38}
\nu^-(R)=\nu^+(R) \ ,
\end{equation}
and
\begin{equation}\label{eq:39}
1-\frac{2M_0}{R}+\alpha f^*_R=e^{-\lambda^+(R)} \ .
\end{equation}
Here $M_0=m(R)$ and $f^*_R$ is the minimal geometric deformation at the boundary of the fluid distribution.
Considering the Israel-Darmois matching condition at the surface~($r=R$), we get second fundamental form that reads 
\begin{equation}\label{eq:40}
G_{\mu\nu} r^\nu \vert_{r \rightarrow R^+}-G_{\mu\nu} r^\nu\vert_{r \rightarrow R^-}=0 \ ,
\end{equation}
where $r^\mu$ is the unit radial vector. Proportionality of $G_{\mu\nu}$ and $T_{\mu\nu}$ implies 
\begin{equation}\label{eq:41}
T_{\mu\nu}^\mathrm{(tot)} r^\nu \vert_{r \rightarrow R^+}-T_{\mu\nu}^\mathrm{(tot)} r^\nu\vert_{r \rightarrow R^-}=0 \ .
\end{equation}
Eq. (\ref{eq:41}) can be rewritten with the help of Eq.~(\ref{eq:12}) in the form 
\begin{equation}\label{eq:42}
(p- \alpha \theta_{1}^{1}) \vert_{r \rightarrow R^+}-(p- \alpha \theta_{1}^{1})\vert_{r \rightarrow R^-}=0 \ .
\end{equation}
As there is no fluid (matter) assumed outside the configuration, we can write 
\begin{equation}\label{eq:43}
p_R-\alpha  (\theta_{1}^{1})_{R}^-=-\alpha  (\theta_{1}^{1})_{R}^+ \ ,
\end{equation}
where $p_R=p^-(R)$.
With the use of Eq.~(\ref{eq:30}) for the inner and the outer geometry, we can write
\begin{equation}\label{eq:44}
p_R+ \alpha \frac{f^{*}_{R}}{k^2}\left(\frac{1}{R^2}+\frac{\nu^{'}_{R}}{R}\right)=\alpha \frac{g^*_R}{k^2}\left[\frac{1}{R^2}+\frac{2M_s}{R^3} \frac{1}{\left(1-\frac{2M_s}{R}\right)}\right] \ ,
\end{equation}
here $g^*_R$ is the deformation of the outer geometry due to the matter term $\theta_{\mu\nu}$, and $M_s$ is the Schwarzschild mass function. 
So, for matching the interior geometry with the exterior one, Eqs~(\ref{eq:38}), (\ref{eq:39}) and (\ref{eq:44}) are necessary and sufficient condition. If the outside geometry is the Schwarzschild vacuum one, Eq.~(\ref{eq:44}) reduces to
\begin{equation}\label{eq:45}
p_R+ \alpha \frac{f^{*}_{R}}{k^2}\left(\frac{1}{R^2}+\frac{\nu^{'}_{R}}{R}\right)=0 \ ,
\end{equation}
which implies that the effective radial pressure at the surface should vanish.

%%%%%%%%%%%%%%%%%%%%%%%%%%%%%%%%%%%%%%%%%%%%%%%%%%%%%%%%%%%%%%%%%%%%%%%%%%%%%%%%%%%%%
\begin{figure*}[t!]
\begin{center}
\includegraphics{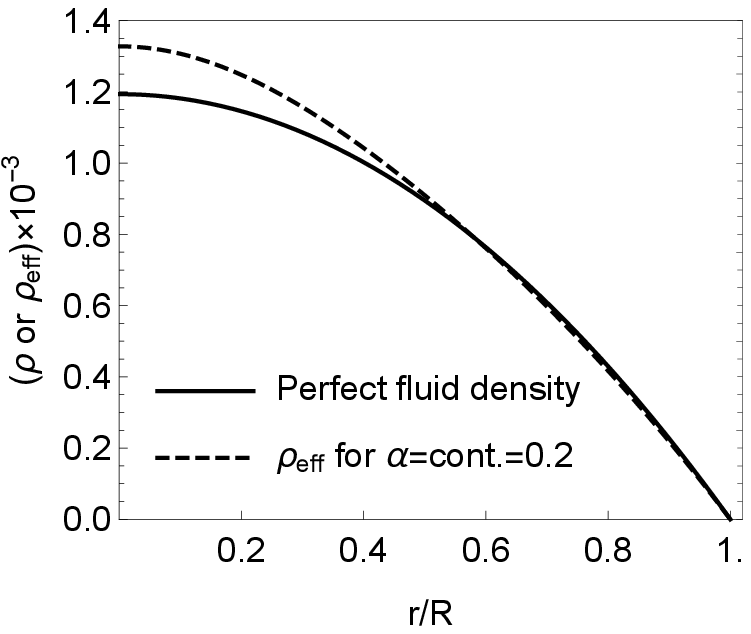}
\hspace{0.7cm}
\includegraphics{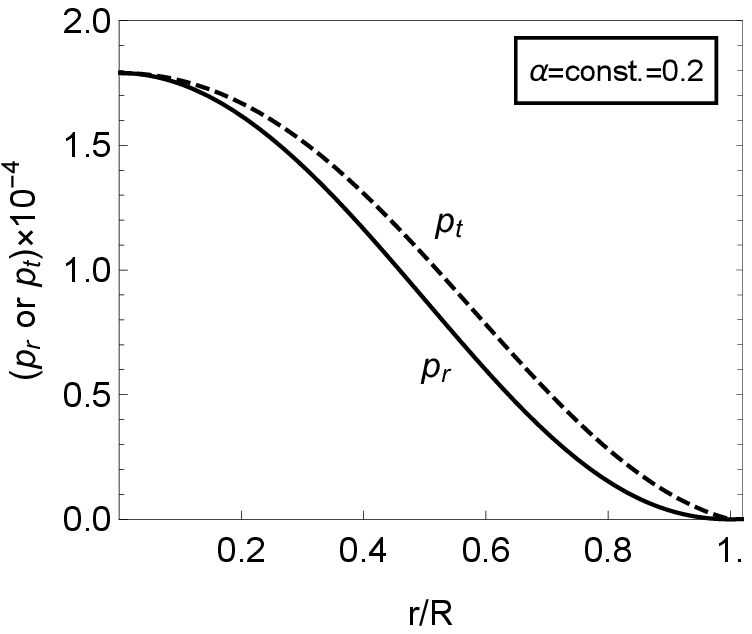}
\end{center}
\caption{Radial dependance of the effective quantities of the new anisotropic solution are shown for the coupling constant $\alpha=0.2$ . The pictures are drawn for $M_0=2$ and $R=10$. \label{effquan}}
\end{figure*}

\section{Interior perfect fluid Tolman VII solution}\label{perfect}

Let us summarize properties of the  well known interior Tolman VII perfect fluid solution \cite{tolman1939static} for which we apply the MGD method. Its metric coefficients are given by
\begin{equation}\label{eq:46}
e^{\nu(r)}=B^2\left[\sin\left(\log\sqrt{\frac{e^{-\gamma(r)/2}+\frac{2r^2}{A^2}-\frac{A^2}{4R^2}}{C}}\right)\right]^2 \ ,
\end{equation}
\begin{equation}\label{eq:47}
e^{-\gamma(r)}=1-\frac{r^2}{R^2}+\frac{4r^4}{A^4} \ ,
\end{equation}
where A,B,C are constants of the solution, yet to be determined. The line element for this solution is described by Eq.~(\ref{eq:19}) with $e^{\epsilon(r)}=e^{\gamma(r)}$. Using Eqs~(\ref{eq:25}) and (\ref{eq:26}), we calculate radial profiles of the pressure and density of the perfect fluid  
\begin{equation}\label{eq:48}
p(r)=\frac{-A^4+4 R^2 r^2+4 R^2 b_{1} A^2 \cot z }{8\pi R^2 A^4} \ ,
\end{equation}
and
\begin{equation}\label{eq:49}
\rho(r)=\frac{-\frac{20 r^2}{A^4}+\frac{3}{R^2}}{8\pi} \ ,
\end{equation}
where $z=\left[\log\sqrt{\frac{\frac{2r^2}{A^2}-\frac{A^2}{4R^2}+ b_{1}}{C}}\right]$, $b_{1}=\sqrt{1+\frac{4 r^4}{A^4}-\frac{r^2}{R^2}}$. The constants A, B and C are calculated according to the matching conditions given by Eqs~(\ref{eq:37}) and (\ref{eq:40}), under assumption of the outside Schwarzschild vacuum solution, and read
\begin{equation}\label{eq:50}
A=\pm \left(\frac{4R^5}{R-2M_0}\right)^\frac{1}{4} \ ,
\end{equation}
\begin{equation}\label{eq:51}
B=\pm \csc  \left[cot^{-1} \left(\frac{M_0 b_{2} b_{3}}{R^7}\right)\right] b_{2} \ ,
\end{equation}
\begin{equation}\label{eq:52}
C=\frac{e^{-2\cot^{-1}\left(\frac{M_0 b_{2} b_{3}}{R^7}\right)}\left(2R^3 b_{2}-4M_0 b_{3}+R b_{3}\right)}{2R^3} \ ,
\end{equation}
where $\frac{M_0}{R} \leq 4/9$ , $b_\mathrm{2}=\sqrt{1-\frac{2M_0}{R}}$ , $b_\mathrm{3}=\left(\frac{R^5}{-2M_0+R}\right)^{\frac{3}{2}}$, and $M_\mathrm{0}=m(R)$ is the total mass of the fluid configuration given by Eq.~(\ref{eq:20}). For the realistic fluid configurations, the perfect fluid density should be positive everywhere within $0<r<R$. For such configurations, the condition $\rho^2 < \frac{3R^2}{5(R-2M_0)}$ has to be satisfied for all values of $r$. The configurations that do not satisfy this condition are not physically acceptable. Examples of the density and pressure radial profiles of the Tolman~VII solution are shown in Fig.~\ref{perquan}\ . We can see they resemble profiles attained for~(neutron) stars and could be thus applied in astrophysical context. For detailed study and application of the Tolman~VII solution see~\cite{tolman1939static}.
%%%%%%%%%%%%%%%%%%%%%%%%%%%%%%%%%%%%%%%%%%%%%%%%%%%%%%%%%%%%%%%%%%%%%%%%%%%%%%%%%

\section{Anisotropic Tolman VII solution by gravitational decoupling}\label{aniso}

\begin{figure}[t!]
\begin{center}
\includegraphics{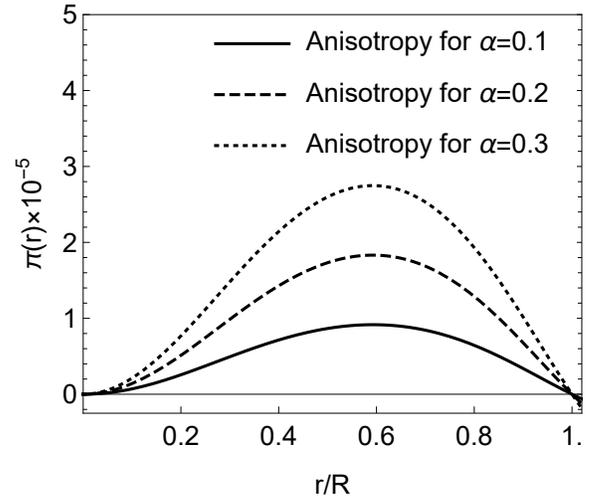}
\end{center}
\caption{In this figure radial dependance of anisotropy is plotted for different values of coupling constant($\alpha$). The pictures are drawn for $M_0=2$ and $R=10$. \label{anisoquan}}
\end{figure}
Now let us turn-on the parameter~$\alpha$ to get an anisotropic solution. According to the matching condition~(Eq.~(\ref{eq:45})), the effective pressure should vanish at the boundary which implies from Eq.~(\ref{eq:12}) that $p_R \sim \alpha (\theta_{1}^{1})_{R}$\ .
If we choose the deformation in the radial component of metric 
\begin{equation}\label{eq:53}
f^{*}=-e^{-\gamma}+\frac{1}{1+r \nu^{'}(r)} \ ,
\end{equation}
it essentially means that 
\begin{equation}\label{eq:54}
\theta_{1}^{1}(r)=p(r) \ .
\end{equation}
Eq.(\ref{eq:54}) is denoted as \lq mimic' constraint by Ovalle~\cite{ovalle2017decoupling}, and it requires to effective pressure to vanish at the boundary.
Using Eqs.~(\ref{eq:22}) and (\ref{eq:48}), we get the radial metric component 
\begin{equation}\label{ea:55}
e^{-\lambda}=e^{-\gamma}+ \alpha \left(\frac{A^2 b_{1}}{A^2 b_{1} + 4 r^2 \cot z }-e^{-\gamma}\right) \ .
\end{equation}
So, Eqs~(\ref{eq:46}) and (\ref{ea:55}) represent the Tolman VII solution being minimally deformed by the gravitational source $\theta_{\mu\nu}$; the original Tolman VII perfect fluid solution can be recovered while $\alpha \rightarrow 0$. 
Now, we have a new anisotropic solution~((\ref{eq:46}) and (\ref{ea:55})). Our task is to match the new solution with the exterior Schwarzschild vacuum metric. Due to the matching conditions~(Eqs. (\ref{eq:38}) and (\ref{eq:39})) we arrive to the relations 
\begin{equation}\label{eq:56}
B^2 \sin^2 z \vert_{r=R}=1-\frac{2M_s}{R} \ ,
\end{equation}
and
\begin{equation}\label{eq:57}
(1-\alpha)e^{-\gamma} \vert_{r=R} + \alpha\left( \frac{A^2 b_{1}}{A^2 b_{1} + 4 r^2 \cot z }\right) \vert_{r=R}=1-\frac{2M_s}{R} \ .
\end{equation}
By using Eq.~(\ref{eq:20}), the above equation gives the Schwarzschild mass($M_s$) due to the relation
\begin{equation}\label{eq:58}
\frac{2M_s}{R}=\frac{2M_0}{R}+\alpha\left(1-\frac{2M_0}{R}\right)- \alpha \frac{A^2 b_{1}}{A^2 b_{1} + 4 r^2 \cot z } \vert_{r=R} \ .
\end{equation}

Considering the Schwarzschild vacuum outside, the second fundamental form~(Eq. (\ref{eq:43})) reads  
\begin{equation}\label{eq:59}
p_R-\alpha  (\theta_{1}^{1})_{R}^-=0 \ ,
\end{equation}
and as a consequence of the \lq mimic' constraint~(Eq.~ (\ref{eq:54})), it reduces to the condition 
\begin{equation}\label{eq:60}
p_R=0 \ .
\end{equation}
Using  Eq. (\ref{eq:48}) this implies for the constant $C$ the relation
\begin{equation}\label{eq:61}
C=\left(-\frac{A^2}{4R^2}+\frac{2R^2}{A^2}+2\frac{R^2}{A^2}\right) e^{-2 \cot^{-1}\left(\frac{A^4}{8R^2}(\frac{1}{R^2}-\frac{4R^2}{A^4}\right)} \ .
\end{equation}
Using the expression for the Schwarzschild mass given in Eq.~(\ref{eq:58}), we obtain from Eq.~(\ref{eq:56})
\begin{equation}\label{eq:62}
B^2 \sin^2 z =(1-\alpha) \left(1-\frac{2M_0}{R}\right)+ \alpha \frac{A^2 b_{1}}{A^2 b_{1} + 4 r^2 \cot z } \vert_{r=R} \ ,
\end{equation}
from which we can determine $B$ while $C$ is given by Eq.~(\ref{eq:61}). Eqs.~(\ref{eq:61}) and (\ref{eq:62}) are necessary and sufficient conditions for matching the anisotropic interior solution with exterior Schwarzschild vacuum.

By using the \lq mimic' constraint~(Eq. (\ref{eq:54})) in Eq.~(\ref{eq:16}), from Eq.~(\ref{eq:48}) we get the radial profile of the radial pressure in the form 
\begin{equation}\label{eq:63}
p_r(r,\alpha)=(1-\alpha)\frac{(-A^4+4 R^2 r^2+4 R^2 b_{1} A^2 \cot z)}{8\pi R A^4} \ .
\end{equation}
The radial profiles of the effective density and the tangential pressure are then given by the relations 
\begin{equation}\label{eq:64}
\rho_\mathrm{eff}(r,\alpha)=\rho(r)+\delta \rho(r,\alpha) \ ,
\end{equation}
and
\begin{equation}\label{eq:65}
p_t(r,\alpha)=p_r(r,\alpha) +\pi(r,\alpha) \ ,
\end{equation}
here $\delta \rho$ is change in density and $\pi(r,\alpha)$ is measure of anisotropy, being defined as 
\begin{equation}\label{eq:66}
\delta \rho=\alpha \frac{Y_1 \cot z+ Y_2 b_1 \cot^2 z -A^2 b_1 (Y_3 + Y_4 \csc^2 z)}{8\pi A^6 R^4 b_1 (A^2 b_1 + 4r^2 \cot z)^2} \ ,
\end{equation}
\begin{equation}\label{eq:67}
\pi(r,\alpha)=\alpha \frac{r^2 \cot z \csc^2 z(Y_5 \cos z^2 + A^2 b_1 Y_5 \sin z^2)}{4 \pi b_1 Y_5 (A^3 b_1 + 4Ar^2 \cot z)^2} \ ,
\end{equation}
where the parameters are determined by the relations\\
$Y_1=4[160r^8 R^4 + A^8(6r^4-8r^2 R^2+3R^4)+A^4(-64r^6R^2+44r^4 R^4)]$ , \\
$Y_2=16 A^2 r^2 R^2[20 r^4 R^2 + A^2 (-3 r^2+R^2)]$ , \\
$Y_3=[-80r^6 A^4-3A^8(r^2-R^2)+4A^4(8r^4 R^2-5 r^2 R^4)]$ , \\
$Y_4=8 A^4 r^2 R^4$ , \\
$Y_5=4r^4 R^2+A^4(-2r^2+3R^2)$ .

The radial profiles of the radial pressure, the tangential pressure and the effective density are shown in Fig.~\ref{effquan}, and the effect of the anisotropy in dependence on radius and the coupling constant~($\alpha$) is shown in the Fig.~\ref{anisoquan}. 

%%%%%%%%%%%%%%%%%%%%%%%%%%%%%%%%%%%%%%%%%%%%%%%%%%%%%%%%%%%%%%%%%%%%%%%%%%%%%%%%%%%%

\section{Conclusions}\label{conclu}

We started from the Tolman VII perfect fluid solution and using the framework of MGD, we get a new anisotropic solution. We demonstrate that the effect of anisotropy is increasing with increasing coupling constant($\alpha$). We also demonstrate that with increasing radius the radial profile of the anisotropy effect increases reaching its maximum value and then it is decreasing to vanish at the edge of the configuration. We calculate the Schwarzschild mass, $M_s$ (from Eq. \ref{eq:58}) which is same as the mass of the perfect fluid stellar distribution, $M_0$. So, in the case considered here, the anisotropy effect is not affecting the mass of the stellar distribution. The reason behind this is evident from the Fig.~\ref{effquan} where we see that the effective density crosses the perfect fluid density profile. The effect of trapping of null geodesics in the anisotropic solution has the same character as in the perfect fluid solution \cite{neary2001tolman}, because the corresponding effective potential of the null geodesics depends on the metric components $g_{tt}$ and $g_{\phi \phi}$ on, which remain unchanged in comparison with the isotropic solution -- only $g_{rr}$ changes in the case considered in our paper. 
%%%%%%%%%%%%%%%%%%%%%%%%%%%%%%%%%%%%%%%%%%%%%%%%%%%%%%%%%%%%%%%%%%%%%%%%
\section*{Acknowledgements}
S.H. and Z.S. would like to acknowledge the institutional support of the Faculty of Philosophy and Science of the Silesian University in Opava, the internal student grant of the Silesian University Grant No. SGS/12/2019 and the Albert Einstein Centre for Gravitation and Astrophysics under the Czech Science Foundation Grant No. 14-37086.
%%%%%%%%%%%%%%%%%%%%%%%%%%%%%%%%%%%%%%%%%%%%%%%%%%%%%%%%%%
\section*{References}
%\clearpage
\bibliographystyle{apsrev4-1}  %% BibTeX style
\bibliography{reference}

\end{document}